\documentclass[
reprint,
superscriptaddress,
aps,
prx,
]{revtex4-2}

\usepackage{graphicx}
\usepackage{dcolumn}
\usepackage{bm}
\usepackage[utf8]{inputenc}
\usepackage{amsmath, amsthm, amssymb, amsfonts}
\usepackage{accents}
\usepackage[usenames,dvipsnames]{xcolor}
\usepackage{tikz}
\usepackage{array}
\usepackage{dsfont}
\usepackage{mathtools}
\usepackage{soul,xcolor}
\usepackage[colorlinks=true,linkcolor=Blue,citecolor=Blue,urlcolor=Blue]{hyperref}
\usepackage[caption=false]{subfig}
\usepackage[capitalise]{cleveref}

\setstcolor{blue}

\begin{document}

\title{Comment on ``Failure of the simultaneous block diagonalization technique applied to complete and cluster synchronization of random networks''}

\author{Yuanzhao Zhang}

\affiliation{Center for Applied Mathematics, Cornell University, Ithaca, New York 14853, USA}
\affiliation{Santa Fe Institute, 1399 Hyde Park Road, Santa Fe, New Mexico 87501, USA}

\begin{abstract}
	In their recent preprint [\href{https://arxiv.org/abs/2108.07893v1}{\color{Blue}{arXiv:2108.07893}}], S. Panahi, N. Amaya, I. Klickstein, G. Novello, and F. Sorrentino tested the simultaneous block diagonalization (SBD) technique on synchronization in random networks and found the dimensionality reduction to be limited. Based on this observation, they claimed the SBD technique to be a failure in generic situations. Here, we show that this is not a failure of the SBD technique. Rather, it is caused by inappropriate choices of network models. SBD provides a unified framework to analyze the stability of synchronization patterns that are not encumbered by symmetry considerations, and it always finds the optimal reduction for any given synchronization pattern and network structure [\href{https://doi.org/10.1137/19M127358X}{SIAM Rev.\ 62,\ 817–836 (2020)}]. The networks considered by Panahi {\it et al}.\ are poor benchmarks for the performance of the SBD technique, as these systems are often intrinsically irreducible, regardless of the method used. Thus, although the results in Panahi {\it et al}.\ are technically valid, their interpretations are misleading and akin to claiming a community detection algorithm to be a failure because it does not find any meaningful communities in Erdős-Rényi networks.
\end{abstract}

\maketitle

In their recent preprint, Panahi {\it et al}.\ \cite{Panahi2021failure} applied the SBD technique to the stability analysis of complete and cluster synchronization states in random networks. 
They found that the variational equation governing the synchronization stability can only be decoupled to a very limited extent for the random networks considered. Based on this observation, they concluded that ``we have set the expectations straight about the reduction that is realistically achievable from application of SBD to the study of complete and cluster synchronization of generic (random) graphs.''

To see why the conclusions of Panahi {\it et al}.\ are misleading, it is helpful to review some recent developments on the stability analysis of cluster synchronization patterns in complex networks.
Cluster synchronization represents dynamical states in which several internally coherent but mutually independent clusters coexist \cite{kaneko1990clustering,belykh2001cluster,pogromsky2002partial,belykh2008cluster,skardal2011cluster,dahms2012cluster,rosin2013control,nicosia2013remote,fu2013topological,orosz2014decomposition,jalan2016cluster,zhang2017incoherence,hart2017experiments,menara2019stability,tang2019master,zhang2020critical,zhang2021mechanism}.
Because nodes in different clusters can adopt distinct dynamics, their stability analyses are significantly more involved than that of complete synchronization.
A breakthrough was made in 2014, in which symmetry-based techniques were introduced to simplify the stability analysis of cluster synchronization patterns derived from group orbital partitions \cite{pecora2014cluster}.
Specifically, it was shown that the variational equation can be optimally decoupled by finding the irreducible representations (IRRs) of the appropriate network symmetry group.
Many important insights on synchronization patterns have since been generated through the applications of the IRR framework and its generalizations \cite{hart2016experimental,cho2017stable,hart2019topological,salova2020decoupled}.

However, there are a few limitations to this elegant framework.
First, IRR is most suited to treat synchronization patterns directly induced by network symmetries.
In order to analyze more general patterns given by equitable or externally equitable partitions \cite{golubitsky2006nonlinear,belykh2011mesoscale,schaub2016graph}, additional steps must be taken to manually incorporate non-symmetry information into the calculations \cite{sorrentino2016complete,siddique2018symmetry}.
Second, IRRs can be expensive to compute, and their computational cost increases with the number of symmetries in the network.
Since the number of symmetries can increase superexponentially with network size $n$, the computation of the IRRs can become intractable already for networks with a few dozens of nodes.

To overcome these limitations, a symmetry-independent framework based on SBD was introduced in 2020 \cite{zhang2020symmetry}. (The special case of complete synchronization, where symmetry is less relevant, was treated in Ref.~\cite{irving2012synchronization}.)
SBD works by finding a finest simultaneous block diagonalization of the matrices in the variational equation, which encode the synchronization pattern and the network structure.
This guarantees that, for any undirected networks, the variational equation is decoupled to the fullest extent possible, largely independent of the specifics of node dynamics and coupling functions \cite{zhang2020symmetry}.
Due to its symmetry-independent nature, the SBD method can treat all cluster synchronization patterns in a unified and direct fashion, regardless of their origin.
Moreover, the SBD method is computationally efficient, which enables the stability analysis of synchronization patterns in large networks with tens of thousands of nodes \footnote{Code for SBD algorithms are available at \url{https://github.com/y-z-zhang/net-sync-sym} and \url{https://github.com/y-z-zhang/SBD}.}.
Finally, the simplicity and flexibility of the SBD method makes it straightforward to incorporate additional complexities into the model, such as nonidentical nodes \cite{zhang2017nonlinearity} and higher-order interactions \cite{zhang2021unified,salova2021cluster,salova2021analyzing}.
Generalizations of the SBD method have been show to provide optimal reduction in directed networks with non-reciprocal interactions \cite{brady2021forget}.

This brief overview of the existing literature should help to put the recent preprint by Panahi {\it et al}\ \cite{Panahi2021failure} into perspective.
What Panahi {\it et al}.\ showed, in its essence, is that very little dimensionality reduction is possible for the synchronization stability problem in the random networks they consider. 
This is not a ``failure'' of the SBD method, which is guaranteed to find the best reduction possible. 
Rather, the lack of reduction is an intrinsic property of the systems chosen by the authors and has little to do with which technique is being applied.
For this reason, it is extremely misleading to proclaim the ``failure of the simultaneous block diagonalization technique'' already in the title, when in fact no other technique---existing or yet to be developed---would be able to find a better reduction for the systems considered.

In our opinion, most random networks are poor benchmarks for the SBD technique. 
A system of coupled ODEs or discrete maps can be decoupled only when there are internal structures that can be exploited, and two or more matrices can have nontrivial common blocks only when they share nontrivial common invariant subspaces.
Thus, aside from complete synchronization in a single-layer network \cite{pecora1998master,lu2006new,nishikawa2006maximum}, the synchronization stability problem in most random networks is not expected to be reducible beyond simply separating the perturbations parallel and transverse to the synchronization manifold. 
This fact is well-known in the synchronization literature---for example, there is usually no dimensionality reduction for clusters that are intertwined \cite{pecora2014cluster}, and the networks considered by Panahi {\it et al}.\ almost always have fully intertwined clusters.
A good analogy here would be community detection in networks---one should not expect to find nontrivial communities in Erdős–Rényi networks, and it would be unfair to say that a community detection algorithm is a failure because it does not find nontrivial communities in such networks.

Which classes of networks would then be more appropriate for benchmarking the performance of methods like IRR or SBD?
The networks must be structured: some examples considered in previous work include networks with a high number of symmetries \cite{pecora2014cluster,zhang2020symmetry} and networks with nonrandom intercluster connections \cite{siddique2018symmetry,zhang2021unified}.
To be fair, it is acknowledged in Panahi {\it et al}.\ that significant reduction can be routinely found through SBD in structured networks. 
However, they consider these cases to be nongeneric.
If there is one thing we learned from the past 20 years of network science research, it is that real-world networks are not random. 
Instead, they are highly structured, and a primary goal of network science has been to understand and leverage the internal organization of network systems \cite{newman2018networks}.
Thus, our difference with Panahi {\it et al}.\ boils down to what networks should be considered ``generic'' or ``interesting'' in the context of cluster synchronization.
However, regardless of one's opinion to this question, SBD remains a general and efficient method for the stability analysis of networked dynamical systems.
The ease of use, guaranteed optimal reduction, and high interpretability of SBD make it an ideal starting point when analyzing any cluster synchronization pattern encountered in nature as well as in the lab.

\bibliography{ref,net_dyn}

\end{document}